\documentclass[twoside,slac_one]{revtex4}
\usepackage{graphicx}
\usepackage{fancyhdr}
\usepackage{amsmath} % American Mathematics Society standards
\usepackage{bm}% bold math
\usepackage{amsxtra}
\usepackage{amssymb}
\usepackage{amsthm}
\usepackage{latexsym}
\usepackage{lscape}
\usepackage{subfigure}

\begin{document}

\title{The Dark Energy Survey Data Management System}


\author{I.Sevilla}
\affiliation{Centro de Investigaciones Energ\'eticas Medioambientales y Tecnol\'ogicas, Av.Complutense 40, 28040 Madrid, Spain}
\author{R.Armstrong, M.Jarvis}
\affiliation{Physics \& Astronomy, University of Pennsylvania, 209 South 33rd St., Philadelphia, PA 19104, USA}
\author{E.Bertin}
\affiliation{Institut d'Astrophysique de Paris, UMR 7095 CNRS, Universit\'e Pierre et Marie Curie, 98 bis boulevard Arago, F-75014 Paris, France}
\author{A.Carlson, S.Desai, J.Mohr}
\affiliation{Department of Physics, Ludwig-Maximilians Universit\"{a}t, Scheinerstr. 1, 81679 M\"{u}nchen, Germany}
\author{G.Daues, M.Gower, R.Gruendl, D.Petravick, T.Tomashek, Y.Yang}
\affiliation{National Center for Supercomputing Applications, University of Illinois, 1205 West Clark Street, Urbana, IL 61801, USA}
\author{W.Hanlon}
\affiliation{Department of Physics, University of Illinois, 1110 West Green Street, Urbana, IL 61801, USA}
\author{R.Kessler}
\affiliation{Department of Astronomy \& Astrophysics, The University of Chicago, 5640 S.Ellis Ave, Chicago, IL 60637, USA}
\affiliation{Kavli Institute for Cosmological Physics, The University of Chicago, 5640 S.Ellis Ave, Chicago, IL 60637, USA}
\author{N.Kuropatkin, H.Lin, J.Marriner, D.Tucker, B.Yanny}
\affiliation{Center for Particle Astrophysics, Fermi National Laboratory, P.O. Box 500, Batavia, IL 60510, USA}
\author{E.Sheldon}
\affiliation{Physics Department, Brookhaven National Laboratory, P.O. Box 5000 Upton, NY 11973, USA}
\author{M.E.C.Swanson}
\affiliation{Harvard-Smithsonian Center for Astrophysics, 60 Garden St, Cambridge, MA 02138, USA}
\author{for the DES Collaboration}

\begin{abstract}
The Dark Energy Survey (DES) is a project with the goal of
building, installing and exploiting a new 74 CCD-camera at
the Blanco telescope, in order to study the nature of cosmic
acceleration. It will cover 5000 square degrees of the
southern hemisphere sky and will record the positions and
shapes of 300 million galaxies up to redshift 1.4. The
survey will be completed using 525 nights during a 5-year
period starting in 2012. About O(1 TB) of raw data will be
produced every night, including science and calibration
images.  The DES data management system has been designed
for the processing, calibration and archiving of these data.
It is being developed by collaborating DES institutions, led
by NCSA. In this contribution, we describe the basic
functions of the system, what kind of scientific codes are
involved and how the Data Challenge process works, to
improve simultaneously the Data Management system algorithms
and the Science Working Group analysis codes.
\end{abstract}

\maketitle

\thispagestyle{fancy}


%%%%%%%%%%%%%%%%%%%%%%%%%%%%%%%%%%
\section{Introduction}
The Dark Energy Survey Data Management (DESDM) is the part
of the Dark Energy Survey project~ \cite{abdalla} that will
transfer, process and distribute the data generated by
survey's camera DECam~\cite{hao}. DESDM is a large, scalable
system led by the National Center for Supercomputing
Applications at the University of Illinois at
Urbana-Champaign (NCSA/UIUC) consisting of:
\begin{enumerate}
\item An archive system for different levels of data.
\item Scientific codes to process raw data.
\item Database to support calibration, provenance and data analyses.
\item Web portals providing process control and easy access
to images and catalogs.
\item Hardware platforms for execution and storage.
\end{enumerate}
As of August 2011, the DESDM team consists of around 20 computing
professionals and physicists from several institutions
around the world, each providing their own expertise to its
development.  In this contribution we describe the basic
design of the DESDM and make particular emphasis on how the
raw data in the form of CCD images is reduced to its final
form as science-ready catalogs. We also present the testing
campaigns that the system is undergoing and the outlook in
the short term, in the context of the DES project. For more
information on other aspects of DESDM, see
also~\cite{mohr}~\cite{kotwani}.
%%%%%%%%%%%%%%%%%%%%%%%%%%%%%%%%%%
\section{Basic functions of the DESDM}
\begin{figure}
\includegraphics[scale=0.35]{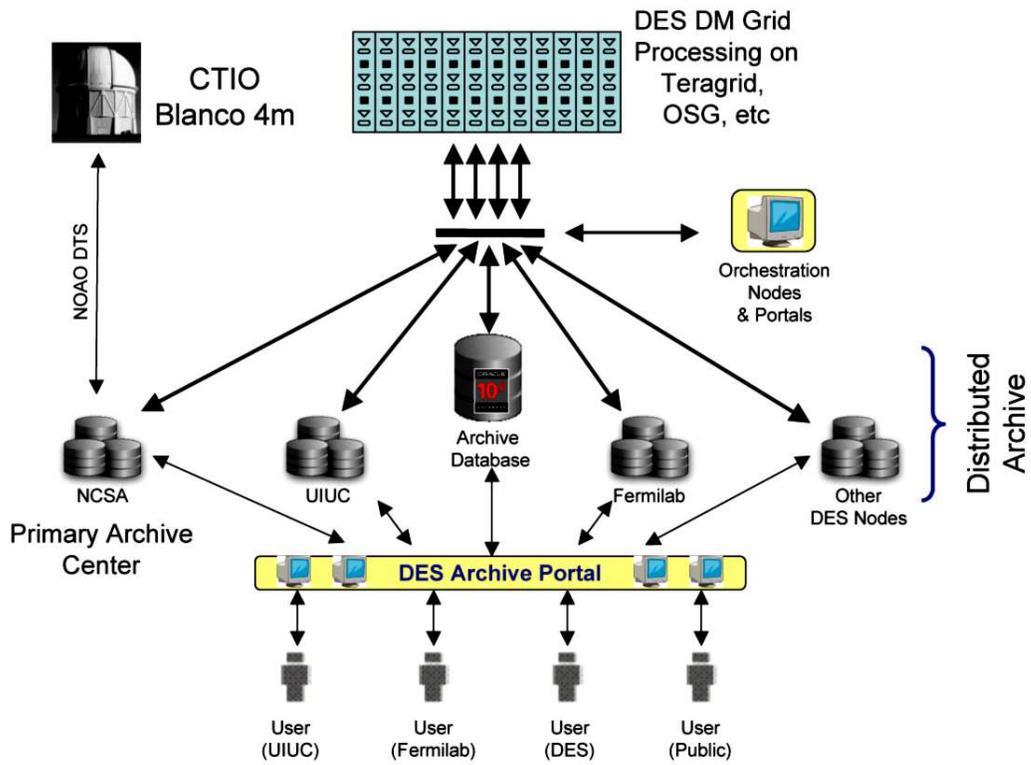}%
\caption{Main data flow and functions of the DES Data
Management system.\label{fig:desdm_diagram}}
\end{figure}

The main functions of the system are schematically
summarized in Fig.~\ref{fig:desdm_diagram}. The design is
driven by the science requirements document of the DES
project, flowed down to the technical level. Additional
requirements include the timely and reliable processing and
the archiving of the data. The functions of the DESDM are the
following: 
\begin{enumerate}
\item\textbf{Transfer.} Raw images must be transferred from
the telescope site at the rate of approximately 100
Mbps. This rate takes into account that 360 science and
associated calibration exposures are produced each night,
totalling $\sim300$ GB of data that have to be transported
in less than 18 hours, to allow for nightly processing and
feedback. Some overhead is included to consider possible
network outages. Recent studies indicate normal data
delivery will occur in near real time.  This step is taken
care of by the National Optical Astronomy Observatory Data
Transport System through a microwave downlink off the
mountain and then by network to NCSA.
\item\textbf{Processing.} As the data arrive at NCSA, it is
ingested into the system, and sent to High Performance
Computing (HPC) resources on XSEDE \cite{xsede} and possibly
the Open Science Grid \cite{osg}. Data are staged using
using GridFTP and GlobusOnline \cite{globus}. The pipeline
containing the parallelization, scientific codes (see
section \ref{sec:algorithms}) and quality assurance is
executed for the night's images.
\item\textbf{Archive and distribution.} Results are then
returned to the NCSA primary archive, and releases
built. Data files in releases are replicated to the
secondary archive at Fermilab and tertiary archives at
collaborator's sites. Data are also released generally to
the collaboration. Catalogs of objects from coadded images
and single epoch images, as well as other meta-data are
served to the collaboration using an NCSA-provided Oracle
RAC system.  After a proprietary period, data will be
released to the community using the same methods. We expect
the system will evolve to use VAO protocols and tools in
time for the public release.
\end{enumerate}
%%%%%%%%%%%%%%%%%%%%%%%%%%%%%%%%%%
\section{Processing the data: from photons to catalogs}
\label{sec:algorithms}
In this section we describe what algorithms are involved in
the reduction of the raw data once the process that sends it
to the HPC platform has been started.
\subsection{Nightly processing}
\begin{figure}
\includegraphics[scale=0.45]{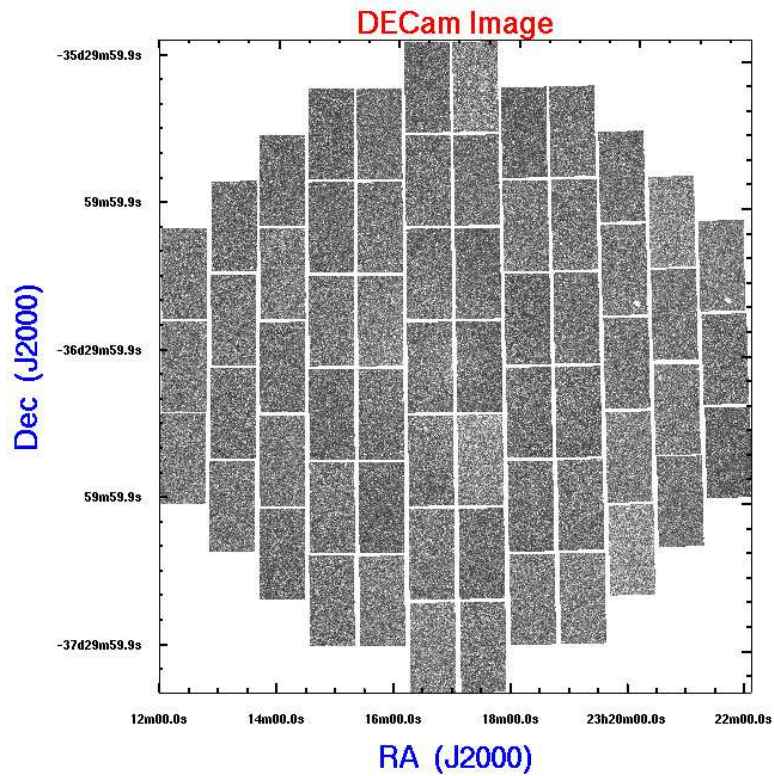}%
\caption{Simulated image showing the appearance of a single
DECam exposure.\label{fig:sample_image}}
\end{figure}

The basic 'event' in an astronomical observation is the
\textit{exposure}. The DECam camera is exposed to the night
sky for $\sim100$ seconds, generating a file of slightly
less than 1 GB in size, which is written in FITS format. An
exposure (see Fig.\ref{fig:sample_image} for an example)
consists of 62 CCD images of a part of the sky, covering
a total solid angle of 3 square degrees, showing multiple
sources and instrumental effects.  Around 300 scientific
images are generated per night, for different pointings,
together with about 60 calibration images. The former have
to be corrected for instrumental effects (detrending) as
well as calibrated for the absolute position (astrometry)
and absolute flux (photometry).
\subsubsection{Image detrending}
This pipeline contains several astronomy modules that
together remove the DECam instrumental signatures. It
includes the following steps:
\begin{itemize}
\item Exposure segmentation and crosstalk correction. The
exposure is divided into 62 CCD images and crosstalk between
them is accounted for, using the crosstalk coefficients
measured from the calibration dataset
\begin{equation}
I_{CCD\_i}(x,y) = \sum_{CCD\_j} \alpha_{ij}I_{CCD\_j}(x,y)
\end{equation}
where $I(x,y)$ stands for the image of a particular CCD and
$\alpha_{ij}$ is the crosstalk coefficient describing the
fraction of image $j$ appearing in image $i$.
\item Image correction. This involves correcting for the
pedestal or bias level of the CCDs, eliminating non-imaging
sections of the exposure, eliminating ghost images from
multiple scattered light in the optics, removing
illumination and fringing effects and correcting for
pixel-to-pixel sensitivity variations
(flat-fielding). Thermal noise from electrons is also
treated but is usually negligible at the camera's working
temperature. In this step additional calibration images
(taken during the same night, every season or during
commissioning) are required and must be part of the job
submitted to the HPC resource.
\end{itemize}
After this process, raw data has been turned into reduced
images with associated maps containing information on bad pixels and the
weights of the image at each particular pixel (inverse variance). 
\subsubsection{Astrometric calibration}
This step requires the identification of positional
standards in the exposure, i.e., stars with very well-known
positions in celestial coordinates. The coordinates of the
brightest sources in the exposure are extracted as well
(using the SExtractor~\cite{sextractor} package) in the
exposures's own reference system in \textit{x}-\textit{y}
coordinates. Knowing an approximate initial solution
(provided by the telescope's control system) a match
of both catalogs can be performed, as well as the fitting of
the best transformation parameters from the image
\textit{x}-\textit{y} system to the celestial reference
system (absolute system in spherical coordinates). This part
is performed by the SCAMP~\cite{scamp}
software. Additionally, the effect of the distortion caused
by the optics towards the edges of the field, has to be
included when calculating the astrometric solution.
\subsubsection{Photometric calibration}
The actual digital counts observed in each pixel of the
detrended image, have to be translated into physical flux
units. In order to make this
conversion, on every night in which conditions are
sufficiently good (moonless, stable skies) specific
calibration star fields will be imaged. For these stars, the
flux is known or can be obtained ($m_{std}$) and this can be
compared to the total light measured in the image
($m_{inst}=-2.5log(counts/sec)$). This relationship is called
the photometric equation and contains several unknowns which
are CCD and filter dependent:
\begin{equation}
m_{inst}-m_{std} = a_{CCD}+b_{CCD}(color-color_0)+kX
\end{equation}
where $a_{CCD}$ is the photometric zeropoint, representing
the normalization value of the image; $b_{CCD}$ is an
instrumental coefficient known as \textit{color term} which takes
into account the shape of the response of the CCD to light
in different filters ($color$); and $k$ is the atmospheric
extinction coefficient which considers the increased
absorption by the atmosphere at the airmass denoted by
$X$ (dependent on sky angle).

Measuring multiple reference stars in the CCDs for different
filters and angles in the sky provides us with enough
equations to solve the system, thus allowing us to find the
photometric solution (values of these constants and their
errors).
%%%%%%%%%%%%%%%%%%%%%%%%%%%%%%%%%%
\subsection{Coaddition}
In order to reach the scientific requirements of DES in
terms of depth, it is necessary to perform a process called
image coaddition. As the name implies, this is the
combination of several overlapping single-epoch images in a
given filter to improve the signal to noise of real sources
in the image. Moreover, if the different images thus
co-added are slightly offset from one another, this
procedure has the added bonus of improving the photometric
calibration with a more robust determination of the
solution, as the photometric solution will incorporate
information from different parts of the sky simultaneously.
Another advantage is the possibility of eliminating
transient effects from the final coadded image such as
satellite trails or cosmic rays, as they will show up in
only one of the images being added and are easily
identifiable.

Prior to the coaddition itself, it is necessary to transform
the flux values in the individual overlapping CCD pixels
into a uniform pixel grid in which the coaddition can be
performed (\textit{remapping}). To do this, artificial
\textit{tiles} in the sky are created, one degree on a side,
and one coadded image per band is produced for every
tile. For each single-epoch image, it is determined to which
tiles it contributes to (using SWarp~\cite{swarp}). The
photometric solution (in particular, the zeropoints) is
re-evaluated for these new images.

There are two main caveats to be pointed out: 
\begin{itemize}
\item the color terms of the CCDs, which are simple to take
into account in single-epoch images, are not easily combined
if they vary across the field. A solution is currently under
implementation;
\item the point spread function (PSF) changes within an image
and from one image to another, due to varying conditions of
the exposure and quality of the sky. Therefore the PSF of the coadded
image is subject to discontinuous jumps.
\end{itemize}

To deal with the second point above, a PSF homogenization
procedure has been developed. The model PSF and its
variation has been computed for each image during nightly
processing using the PSFEx~\cite{psfex} package. We
define the target PSF to be used as a circular Moffat \cite{moffat} function 
with a Full Width at Half-Maximum which is
the median of the seeing distribution one in the
whole set of images contributing to the coadd. Going back to
the individual component images with PSF $\Psi$, we find the
kernel $\kappa$ which minimizes the difference with respect
to this median target PSF $\Psi_{median}$.
\begin{equation}
\chi^2 = \lvert \Psi - \sum_{l}Y_l(x_i)\kappa_l\ast\Psi_{median} \rvert^2
\end{equation}
where $Y_l$ are the elements of a polynomial basis in $x-y$.
The kernel elements are stored and during the
homogenization process, they are recovered to convolve with
the image. Once every image has been treated this way, the
coaddition will not introduce any discontinuities from the
PSF variations.

 Image coaddition takes place off-season due to its increased CPU
requirements with respect to the nightly processing. Coadd construction is to be carried out multiple
times depending on the specific needs of each Science
Working Group, which have different requirements in the
balance between depth and source morphology. For instance,
the Weak Lensing group would be interested in no coaddition
at all given their stringent requirements on the shape
measurement (though they will benefit from multiple imaging
of the same objects, improving the solution of the extracted
shear, see section~\ref{sec:wl}).

%%%%%%%%%%%%%%%%%%%%%%%%%%%%%%%%%%
\subsection{Cataloging}
The final step in the processing consists in transforming
the coadded images in different filters into single object
entries in a catalog. DESDM uses the SExtractor software in
this step too, running over a master image which uses coadd
images in all filters to detect where the sources are, and
then extracting the relevant information from single-filter
coadded images. An object is identified as such when the
convolution of the PSF with the image is above a certain
local background estimation (see \cite{sextractor} for
details). In this step, it is important to approach the
deblending of sources. This is currently being done by
producing several isophotal layers for each object and at
each layer where two light-'islands' join making the
decision on whether to merge them into the same object using
as a criterion the relative integrated intensity between the
branch and full object.

The type of information extracted in this step is
positional, photometric and morphological, besides
identification and other bookkeeping variables.

Concerning the astrometry, the barycenter for each object is
derived using several estimators.  The most reliable one
makes use of an iterative calculation through a Gaussian
window. This is transformed to celestial coordinates using
the astrometric solution found during nightly processing.

Photometry is measured using several methods:
\begin{itemize}
\item simple flux counting inside a fixed circular aperture for
several apertures (in the arcseconds range); 
\item using an elliptical aperture adjusted to the
morphological properties of the object. From the second
order moments of the object, we would find the elongation
and orientation of the ellipse representing it. The ellipse
scaling factor is derived from the first order moment of the
radial distribution \cite{kron};
\item using a fit to the measured PSF shape (suitable for
stars);
\item using a one- or two- component model convolved with a
local model of the PSF (exponential and/or spheroidal,
suitable for galaxies).
\end{itemize}

Among the morphological measurements, currently there are
two variables appraising the deviation of the object from a
point-like shape, therefore providing a star-galaxy
separation handle. One of them relies on a previously
trained neural network and is the well-known stellarity
parameter CLASS\_STAR from SExtractor. The other arises from
the calculation of a discriminant function measuring the
deviation of the image from the PSF shape.

Most of the variables in the catalogs have their corresponding errors and
quality flags to guide the selection of sources for
analysis.

\subsection{Additional Pipelines}
Additional columns are included in the final catalog, and
these are produced in pipelines running over images and
catalogs, to produce the final coadded catalogs. Two of
these are briefly described in the following sections. An
additional catalog is built to identify transient objects,
as described in section~\ref{sec:diff_imaging}.
\subsubsection{Photometric Redshifts}
\begin{figure}
\includegraphics[scale=0.20]{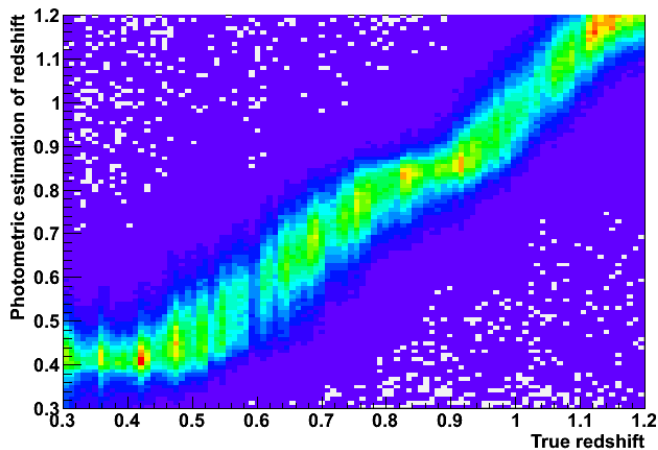}%
\caption{Color plot for photometric redshift estimation vs true redshift for the galaxy catalog generated in the latest Data Challenge\label{fig:photoz}}
\end{figure}

Redshifts in the DES will be estimated from the photometric
information of the objects in different
filters. Traditionally, the approach to this has been to
either find the best fit of the spectral energy distribution
to a collection of templates for different types of galaxies
at different redshifts, or use a neural network trained with
spectroscopic information (see~\cite{bpz} and ~\cite{oyaizu}
for examples of both). Currently the default estimation in
DESDM uses the latter method, using ten input magnitudes
(fixed circular and automatic elliptical apertures,
described above, for five filters each). In
Fig.~\ref{fig:photoz} a comparison of photometric redshifts
versus spectroscopic (true) ones is shown, as obtained in
one of the latest Data Challenges (section
\ref{sec:dcs}). There is a large code comparison project
within a specific photo-z Science Working Group.
\subsubsection{Weak Lensing Pipeline}
\label{sec:wl}
The Weak Lensing probe will require measuring the distortion
in the shapes of galaxies to extract the shear produced by
gravity from the intervening matter between the source and
observer. It will require a very precise measurement of the
local PSF shape independently of the standard determination
used in the pipeline. This is done using bright isolated
stars and additional instrument data.  The PSF is
interpolated with a polynomial throughout the image, and it
is deconvolved from each galaxy. The shear will be extracted
using all the images where the object shows up (in multiple
filters) to provide a more robust determination.
\subsubsection{Difference Imaging}
\label{sec:diff_imaging}
The DES will contain two modes of operation: the survey mode
in which a wide area of the sky will be scanned several
times for each filter over the course of the five-year
survey; and a 'time-domain' survey where particular regions
will be observed repeatedly on short time-scales ($\sim$
weekly) in search of transient phenomena (from which
supernovae of type Ia have to be identified).

From every reduced image coming from the nightly processing
of these regions, a template is extracted from the coadd
corresponding to that position. This template is subtracted
from the reduced image, and all objects remaining above a
certain threshold are cataloged. 
\subsubsection{Survey Mask Generation}
The DESDM also plans to automate the generation of a
polygon-based survey mask which would encode information on
which single epoch images contributed to the coadded image
at a given point and track the coadded magnitude depth,
coadded color terms, regions blocked by saturated stars and
any other information which is relevant to be included as a
survey map rather than in a source-by-source basis. The mask
is generated offline using the Mangle~\cite{mangle}
software, which is currently being incorporated to the
pipeline.
%%%%%%%%%%%%%%%%%%%%%%%%%%%%%%%%%%
\section{Testing the Pipelines}
\label{sec:dcs}
\begin{figure}
\centering
\subfigure[Photometry test from one of the Data Challenges. In
the case of the photometry, the requirement is that the bias
of the measurement with respect to truth values must stay
below 0.02 up to the magnitude limit of the band
($\sim24$).] {
\includegraphics[scale=0.20]{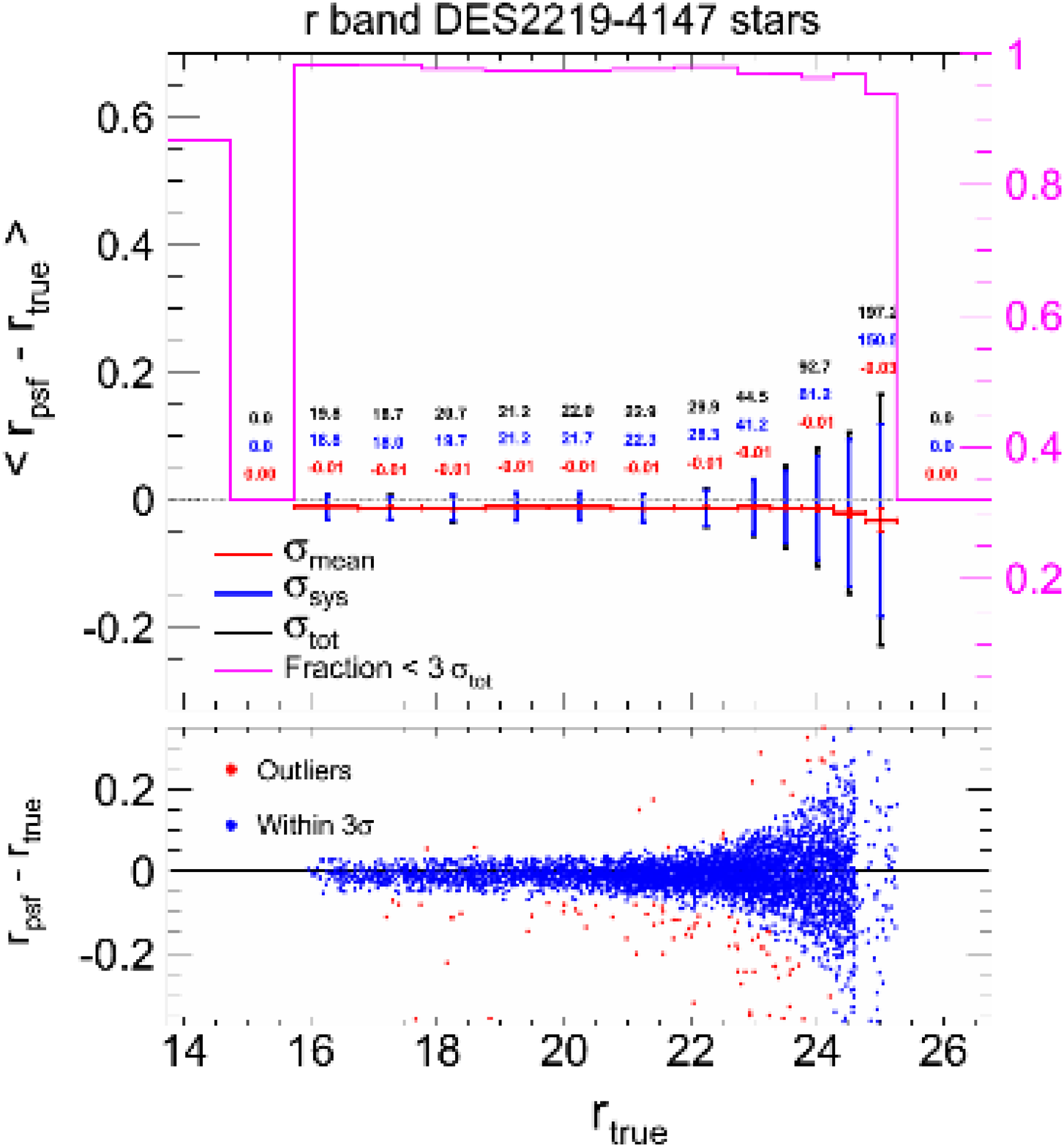}%
\label{fig:photometry}}
\subfigure[Astrometry test from one of the Data
Challenges. For the astrometry, the Gaussian fit must have
$\sigma<100$ milliarcseconds.]{
\includegraphics[scale=0.50]{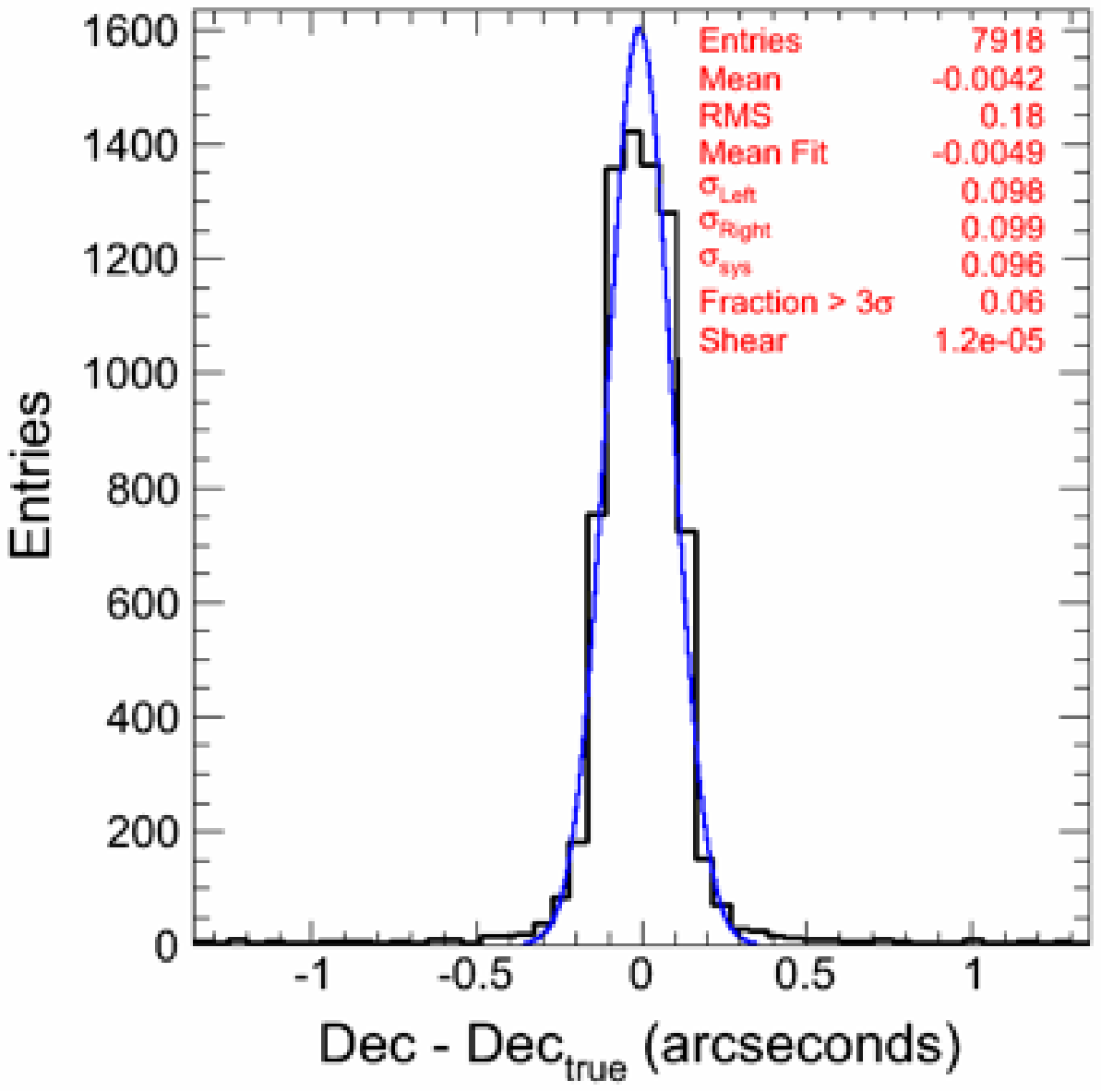}%
\label{fig:astrometry}}
\caption{Sample QA results for the latest Data Challenge \label{fig:qa_sample}}
\end{figure}

During the execution of the pipelines, in-built Quality
Assurance modules make sure that the images and catalogs are
good enough for scientific analysis and diagnose possible
problems. Some sample results are shown in
Fig.\ref{fig:photometry} and Fig.\ref{fig:astrometry}.
 
A broader approach to the testing and validation of the
DESDM is the Data Challenge (DC) process, in which a small
sample of the survey is simulated in detail and fed to
the pipelines to generate realistic images and catalogs. The
scale for the sample ranges from a single 0.6 square degree
tile to 200 square degrees corresponding to about 10 nights
of observations. It starts with the creation of galaxy
catalogs stemming from an N-body simulation\cite{busha} and detailed
models of the Milky Way Galaxy for the star component \cite{trilegal}. These
are merged and fed to an image simulator which includes
atmospheric and instrumental effects. The resulting images
serve as inputs for the DESDM, as if they had been really
observed at the telescope. This whole process is a joint
effort of the Stanford, Brazil and Barcelona teams, for the catalog
side, and Fermilab, for the image simulation aspect. The resulting
catalogs are then examined by members of the DESDM and
Science Working Groups.

Several of these DCs have taken place during the project's
lifetime. Currently the testing approach consists on large
Data Challenges every 6 months to verify the compliance with
science requirements, while maintaining a 2-week cycle with
the generation of small samples for quick feedback to the
developers.

%\begin{figure}
%\includegraphics{dc_results.eps}%
%\caption{Photometry test from one of the Data Challenges. In
%the case of the photometry, the requirement is that the bias
%of the measurement with respect to truth values must stay
%below 0.02 up to the magnitude limit of the band
%($\sim24$).\label{fig:photometry}}
%\end{figure}
%\begin{figure}
%\includegraphics[scale=0.50]{astrometry.eps}%
%\caption{Astrometry test from one of the Data
%Challenges. For the astrometry, the Gaussian fit must have
%$\sigma<100$ milliarcseconds.\label{fig:astrometry}}
%\end{figure}

In addition, the system is being used on real images of the
Blanco Cosmology Survey~\cite{blanco} which are proving to
be very valuable for the development of the algorithms. 

%%%%%%%%%%%%%%%%%%%%%%%%%%%%%%%%%%
\section{Outlook}
The DESDM system has been fully developed and all elements
are in place to process the large dataset that will be
generated by the DES camera. 

Current testing results point to requirements being met in
terms of astrometry, photometry and depth. Others such as
completeness and star-galaxy separation are compliant up to
shallower magnitudes though improvements are in the works.

In parallel, a Community Pipeline is being developed using
essentially the same structure and algorithms but with the
goal of addressing non-DES user needs, given that two-thirds
of the year the instrument will be available to the general
astronomical community with access to NOAO facilities.

The system will undergo a final acceptance test in the next months before
scientific operations start by Fall 2012.

%%%%%%%%%%%%%%%%%%%%%%%%%%%%%%%%%%
\begin{acknowledgments}
MS was supported by the National Science Foundation under  Award No.
AST-0901965.

Funding for the DES Projects has been provided by the
U.S. Department of Energy, the U.S. National Science
Foundation, the Ministerio of Ciencia y Educaci\'{o}n of
Spain, the Science and Technology Facilities Council of the
United Kingdom, the Higher Education Funding Council for
England, the National Center for Supercomputing Applications
at the University of Illinois at Urbana-Champaign, the Kavli
Institute for Cosmological Physics at the University of
Chicago, Financiadora de Estudos e Projetos,
Funda\c{c}\~{a}o Carlos Chagas Filho de Amparo \`{a}
Pesquisa do Estado do Rio de Janeiro, Conselho Nacional de
Desenvolvimento Cient\'{i}fico e Tecnol\'{o}gico and the
Minist\'{e}rio da Ci\^{e}ncia e Tecnologia, the Deutsche
Forschungsgemeinschaft and the Collaborating Institutions in
the Dark Energy Survey.

The Collaborating Institutions are Argonne National
Laboratories, the University of California at Santa Cruz,
the University of Cambridge, Centro de Investigaciones
Energ\'{e}ticas, Medioambientales y Tecnol\'{o}gicas -
Madrid, the University of Chicago, University College
London, DES-Brazil, Fermilab, the University of Edinburgh,
the University of Illinois at Urbana-Champaign, the Institut
de Ci\`{e}ncies de l'Espai (IEEC/CSIC), the Institut de
F\'{i}sica d'Altes Energies, the Lawrence Berkeley National
Laboratory, Ludwig-Maximilians Universit\"{a}t and the
associated Excellence Cluster Universe, the University of
Michigan, the National Optical Astronomy Observatory, the
University of Nottingham, the Ohio State University, the
University of Pennsylvania, the University of Portsmouth,
Universidade Federal do Rio Grande do Sul, SLAC, Stanford
University, the University of Sussex and Texas A\&M
University.
\end{acknowledgments}
\bigskip % extra skip inserted

\end{document}